# Evidence of local softening in glassy poly(vinyl alcohol)/poly(vinyl pyrrolidone) (1/1, w/w) nano-graphene platelets composites

Eirini Kolonelou [1], Anthony N. Papathanassiou [1, *] and Elias Sakellis [1,2]

[1] National and Kapodistrian University of Athens, Physics Department. Panepistimiopolis, GR 15784 Zografos, Athens, Greece

[2] National Center of Natural Sciences Demokritos, Institute of Nanomaterials and Nanotechology, Aghia Paraskevi, Athens, Greece

**ABSTRACT**

Complex permittivity studies on glassy poly(vinyl alcohol)/poly(vinyl pyrrolidone) (1/1) at proper pressure-temperature condition, provide the activation volume and energy values for both dc conductivity and β-relaxation. The temperature dependence of the activation volume, which signatures volume-fluctuations accompanying a dynamic process, maximizes at temperature near the glass transition temperature of neat polyvinyl alcohol. The phenomenon is interpreted by a local softening of polyvinyl alcohol domains, while the blend remains in its glassy state. The scenario is also supported by the dependence of the activation energy upon pressure. Dispersed nano-graphene platelets at volume fractions around the critical conductivity percolation threshold are comparatively studied, so as to determine whether local softening affects the formation of a conductivity percolation network.

1. **Introduction**

Polyvinyl alcohol (PVA) is a water-soluble synthetic polymer, which has excellent film forming, emulsifying and adhesive properties [1]. It has high tensile strength and flexibility comparable to those of human tissues. PVA finds its wide application in food, cosmetic, internal wall coating, plasterwork and joint sealing due to its favorable properties of weather resistance, waterproof, non-swell with water, non-embrittlement, non-poison, tastelessness, and low cost. [2]. Polyvinyl pyrrolidone (PVP) is water-soluble optically transparent polymer. [3, 4] PVP is used in personal care products and adhesives that must be moistened. PVP is soluble in water and other polar solvents, such as alcohols. When dry, it is a light flaky hygroscopic powder, readily absorbing up to 40% of its weight in atmospheric humidity. In solution, it has excellent wetting properties and readily forms films. This makes it good as a coating or an additive to coatings. [5]. Graphene is an allotrope of carbon consisting of a single layer of carbon atoms arranged in an hexagonal lattice. [6, 7]. It is the strongest material ever tested, efficiently conducts heat and electricity and is nearly transparent. Graphene nanoribbons, graphene nanoplatelets, and nano–onions are believed to be non-toxic at concentrations up to 50 μg/ml. These nanoparticles do not alter the differentiation of human bone marrow stem cells towards osteoblasts (bone) or adipocytes (fat) suggesting that, at low fractions, graphene nanoparticles are safe for biomedical applications.[8].

The polymer matrix studied in the present work is *(PVA/PVP):(1/1)*. According to the literature, the glass transition temperatures of PVA and PVP are: $T_{g,PVA}$=358 K, while for PVP $T_{g,PVP}$=441 K, respectively. The glass transition temperature of PVA – PVP: 1-1 can be evaluated through the Fox formula: $T_g^{-1} = w_{PVA}T_{g,PVA}^{-1} + w_{PVP}T_{g,PVP}^{-1}$, where $w_{PVA}$, $w_{PVA}$ denote the mass fractions of PVA and PVP, respectively [9]. Setting $w_{PVA}$=$w_{PVP}$ = 0.5, thus, $T_g$=386 K. In fact, the transition from the glassy state to the rubber one is not abrupt, but spans over a broad temperature region around $T_g$, which is labeling a mean dominant temperature signifying the transition. The Fox equation can only be used for miscible polymer systems (that is the case for PVA/PVP blends) [10]. Nano-graphene platelets (NGPs) were the conductive phase dispersed into (PVA/PVP):(1/1). The electric charge flow in (PVA/PVP):(1/1) and composites loaded with different mass fractions of NGPs is explored at various spatiotemporal scales and isothermal and isobaric conditions below $T_g$, by employing Broadband Dielectric Spectroscopy (BDS). The blend has a glass transition temperature $T_{g,PVA-PVP}$= 386K (which we label as a global one of the blend). The scope of the present work is to examine whether if local softening in a vitreous glassy network of end occurs, by monitoring the activation volume or dc-conductivity and relaxation, respectively, which is a fluctuation volume linked to the correlation of a charge flow event with its immediate environment. BDS at various pressures yields the evaluation of the activation volume related to a dynamic process, which is defined thermodynamically as $\upsilon^{act,i} \equiv \left(\frac{\partial g^{act,i}}{\partial P}\right)_T$ [11], where the index i denotes the process (such as dc conductivity or relaxation) and $g^{act,i}$ is the Gibbs free energy for this process. The activation volume captures information about the elasticity of the material, structural rearrangement, the type of the mobile charge and the nature of a flow event (such as classical migration or thermally assisted quantum tunneling) [12, 13]. Mobile electric charge, at different spatio-temporal scales, can prove local or global changes of the degrees of freedom of a matrix through the study of the corresponding activation volume. For this reason, BDS – which explores different spatio-temporal scales - was applied at isothermal and isostatic conditions below $T_g$. The concept of polyvinyl alcohol domains softens while the blend remains globally glassy is, to the best of our knowledge, novel and can find applications, such as two-stage drug delivery through local and global softening of the polymer matrix.

## 2. Experimental technique

Equal masses of PVA and PVP powders were dissolved separately into hot doubly distilled water by continuous stirring. The aqueous solutions merged together. NGP powder was dissolved into water at 353 K and ultra-sonicated in a heat bath for about two hours. Afterwards, the polymer and water solutions were mixed, and stirred prior to drop-casting on a clean Teflon surface. After 48 hours drying at ambient temperature and humidity, free standing solid specimens of about 1 mm thickness were obtained. SEM microscopy ensured a homogeneous distribution of the dispersed NGPs [14]. A Novocontrol High Pressure BDS system. Dielectric measurements at temperatures above the ambient one and pressure less than 3 kbars, in the frequency range from $10^{-2}$ to $10^6$ Hz. Samples of typical surface area of 1 cm$^2$ and thickness of 1 mm were placed in the pressure vessel following the methodology published earlier [15 - 17]. The opposite parallel surfaces of the free-standing specimens were silver pasted to ensure good electrical contact with the measuring electrodes. Complex permittivity measurements were collected from 1 mHz to 1 MHz with a Solartron SI 1260 Gain-Phase Frequency Response Analyzer, equipped with a Broadband Dielectric Converter (BDC, Novocontrol). Data acquisition was monitored through the WinDeta (Novocontrol) software [18, 19]. The formalism of the complex permittivity ε* and the electric modulus M* as a function of frequency f were employed to study the effect of temperature, pressure and composition on the dielectric properties of the composites.

## 3. Results and discussion

The formalisms of the complex permittivity $\varepsilon^*(f) = \varepsilon'(f) + i\varepsilon''(f)$, where $i^2 = -1$ and f is the frequency of an externally applied harmonic field, M* from the frequency, and the complex electric modulus $M^* = \frac{1}{\varepsilon^*} = M' + iM''$ [20] have both been used to study dc-conductivity, while ε*(f) additionally provides information about the β-relaxation. The mechanisms of dielectric relaxation and electrical conductivity of (PVA/PVP):(1/1) and its composites with dispersed NGPs loading up to 0.3 w-w %, were investigated at temperatures ranging from 293 K to 393 K (whereas the blend is in its glassy state) and hydrostatic pressures up to 2500 bar. A paradigm of the effect of pressure and temperature on $\varepsilon''(f)$ is depicted in Figure 1. As can be seen in Figure 2, the spectra consists of a dc-conductivity straight line and, at high frequencies, of underlying relaxations, appearing - depending on the degree of masking - as knees or peaks. The permittivity data were analyzed by fitting the following equation:

$$\varepsilon''(f) = \frac{\sigma_{dc}}{\varepsilon_0 2\pi f^n} + \frac{\Delta\varepsilon}{\left(1+(f/f_0)^a\right)^b}$$

(1)

where $\sigma_{dc}$ denotes the dc-conductivity, $n$ is a fractional exponent ($n \leq 1$) which is usually close to 1, $\Delta\varepsilon$ is the intensity of a relaxation mechanism, $a$ and $b$ are fractional exponents and $f_0$ is a parameter that coincides with the peak maximum frequency when b=1. Relaxation peaks are assumed to obey the Harviliak - Negami (H-N) model [21] where a and b are symmetric and asymmetric broadening parameters, respectively. For a Debye peak, the parameters amount to unity (a=b=1). A typical fitting of eq. (1) to the data points is depicted in Figure 2, where the spectrum is decomposed to a dc-component and a relaxation peak, which, as being traced in pure (PVA/PVP):(1/1) below the glass transition is attributed to the β-relaxation of the macromolecules. Its position in the frequency domain is in agreement with that reported in the literature [22, 23], as well. To ensure an further accurate determination of the dc conductivity from BDS data, by suppressing the role of undesirable low frequency electrode polarization employed the electric modulus function, defined as $M^*(f)=1/\varepsilon^*(f)$. The temperature and pressure evolution of $\sigma_{dc}$ was alternatively monitored through the shift of the dc-conductivity peak in the imaginary part of the electric modulus: $M''(f) = M_s \frac{f\tau_\sigma}{1+(f\tau_\sigma)^2}$. In the modulus representation, the dc-conductivity line appearing in eq. (1), transforms into a conductivity peak, maximizing at $f_{o,M''}$. A plot of $M''(f)$ against logf yields a Debye-like peak with a maximum at $f_{0,\sigma} = \sigma_0/(2\pi\varepsilon_\infty)$, where $\varepsilon_\infty = \varepsilon'(f \to \infty)$.

Concerning the electrical properties of the specimens, the transition from insulator to semiconductor (IST) occurs around 0.1 % w-w NGP at ambient pressure, and it is practically independent upon temperature. We stress that the IST is related with the formation of a conductivity percolation network as the volume fraction of NGPs increases and is substantially different than the glass transition of the polymers, as can be seen in the Figure 3, where the ambient pressure values of $\log\sigma_{dc}$ are plotted against mass fraction of the NGP. At 1 bar (Figure 3), the percolation threshold seems insensitive, compared to the ambient pressure critical point; however, at elevated temperatures, the percolation threshold moves toward lower concentrations (i.e., below 0.1%), namely when temperature is equal or larger than 373 K. Pressure brings NGP conducting islands close together and the enhanced phonon-assistance of quantum penetration of the polymers by electrons. We note that, in insulating materials with dispersed conducting articles, whereas electrons can tunnel through the insulating barrier, the insulator to conductor transition is smother compared with the sharp one predicted for composites with ideally non-nteracting phases.

The activation volume for dc electrical conductivity, obtained from equation [11]:

$$v^{act,dc} = -2.3 k_\text{B} \left(\frac{\partial \log \sigma_{dc}}{\partial \text{P}}\right)_\text{T} \qquad (2)$$

where $k_\text{B}$ denotes the Boltzmann's constant, is depicted for the polymer matrix and nano-composites of different loading, around the percolation threshold.

All data sets depicted in Figure 4, where $v^{act,dc}$ is plotted against temperature exhibit a maximum around 343 K. Moreover, the activation volume values have an increasing monotony upon NGP fraction, close or below the percolation limit. Above the critical point (i.e., for for 0.3 w-w %vNGP) the activation volume values drop to about one third of those obtained for lower NGP loadings..The change of monotony in $v^{act,dc}(T)$ can be explained as follows: The activation volume is a measure of the difficulty for mass (charge) transport and is interconnected with volume fluctuation per mass flow event [13]. In Figure 4, below the glass transition temperature of PVA ($T_g$=358 K), $v^{act,dc}(T)$ increases monotonically; A local softening above the glass transition temperature of PVA, provides flexible PVA domains; electric charge penetrating such rubber regions induce volume fluctuation effects as a result of easier macromolecular rearrangement in the rubber state, compared to the glassy one. Subsequently, above the glass transition temperature of PVA (and below the global glass transition of the polymer blend), the measured activation volume captures contributions from the glassy state and rubber PVA domains: the latter has a gradually pronounced effect on increasing temperature, yielding in a monotonic decrease of $v^{act,dc}(T)$ above $T_{g,\,PVA}$. According to the above scenario, electric charge transport senses *local* structural softening on increasing temperature, significantly below the glass transition temperature of the whole system ((PVA/PVP):(1/1). $T_g$=386 K, where the system transits *globally* to the rubber state. Transferring electric charges can prove local softening of PVA domains, while the entire system withstands the global glassy features. The material resembles that of a hard porous matrix with encapsulated dispersed domains of PVA, which, depending on temperature, can either be glassy or rubber-like. At higher temperature, reaching $T_g$=386 K, the whole system becomes unstable and undergoes a *global* transition from glassy to rubber state. If the system is loaded with NGP mass fraction beyond the conductivity percolation threshold (e.g., for 0.3 w-w % NGP), electric charge is making better use of the network provided by NGPs, the volume distortion induced is suppressed and, consequently, the activation volume values are lower than those obtained for NGP fraction close or below the critical one (see star data points in Figure 4).

Figure 5 indicates that the dc-conductivity activation energy reduces upon NGP mass fraction, at ambient pressure. The decrease of the activation energy probably results from the ease of macroscopic electric charge transport, as the composites becomes rich in conducting islands and available percolation pathways. At higher pressure, the monotony of: the activation energy changes for NGP fractions lower than that of the percolation threshold. The maximum in the activation energy values is likely to occur due to the competition between two competing phenomena: Firstly, the presence of NGPs into the polymer matrix, might slow down or constrain the dynamics of macromolecules. Secondly, a closer packing of the polymer network is dictated by increased hydrostatic compression, relatively to the ambient pressure state. The combination of these contributions yields an increase of the activation energy at low NGP fractions, while, at higher NGP loading, a dense network of conducting NGPs is attained, facilitating dc-conductivity and lowering of the corresponding effective activation energy values.

A *local* softening, when approaching $T_{g,PVA}$, should affect - apart from the *macroscopic* dc-conductivity – *local* relaxations. The activation volume $\upsilon^{act,\beta}(T)$ for the $\beta$ sub-$T_g$ relaxation is commonly obtained by [11]:

$$\upsilon^{act,\beta} = -2.3 k_B T \left( \frac{\partial \log(f_o)}{\partial P} \right)_T \tag{3}$$

is a measure of the retardation of relaxation on compression. The isothermal pressure evolution of $\log f_0$ (Figure 6) obeys Eq. (3) yielding, $\upsilon^{act,\beta}$. In Figure 7, $\upsilon^{act,\beta}$ is plotted as a function of temperature for different NGP loadings the monotony of. $\upsilon^{act,\beta}(T)$ is independent upon temperature. Local segmental motion is facilitated upon temperature for PVA/PVP and composites loaded with less than 0.3 wt% NGP . Above the percolation threshold an opposite behavior is observed.

The temperature dependence of β-relaxation in polymers is usually described by an Arrhenius law, which yields the estimation of the activation energy $E^{act,\beta} = -2.3 k_B \left( \frac{\partial \log f_o}{\partial T^{-1}} \right)_P$ [11]. In Figure 8, $E^{act,\beta}$ is plotted as a function of NGP mass fraction. The inset diagram of Figure 8, shows the $\log f_0(1/T)$ data points can be fitted by a straight line, ensuring an Arrhenius dependence. $E^{act,\beta}$ values get suppressed when (PVA/PVP):(1/1) and NGPs composites undergo hydrostatic compression (Figure 8). To a first approximation, pressure reduces the volume of the specimen and, therefore, the density of NGPs augments. Accordingly, by increasing either the volume fraction of NGPs or pressure, the activation energy of β-relaxation gets reduced. Pressure restricts the volume available for segmental motions. Subsequently, less work is required for shorter-scale local motions. In other words, decrease of the activation energy is correlated to the mechanical work expensed per relax or for the relaxation of segmental parts of the macromolecules.

**4. Conclusions**

In conclusion, complex permittivity studies on poly(vinyl alcohol)/poly(vinyl pyrrolidone) (1/1) at proper pressure-temperature conditions below the glass transition, provide the activation volume and energy values for dc conductivity and β-relaxation. The temperature dependence of the activation volume, a signature of volume-fluctuations accompanying a dynamic process, senses softening of the glassy blend, at a temperature which practically coincides with the glass transition temperature of the neat polyvinyl alcohol. This scenario is also supported from the pressure variation of the activation energy values. Accordingly, activation volume and energy for β-relaxation sense local softening while the blend remains globally glassy. Furthermore, by dispersing nano-graphene platelets at volume fractions around the critical conductivity percolation threshold, we observed that local softening does not significantly affects the formation of a percolation network.

**Acknowledgments'**

The authors are grateful to Evangelia Roumelioti for experimental contribution and partial preliminary analyses of the raw data.

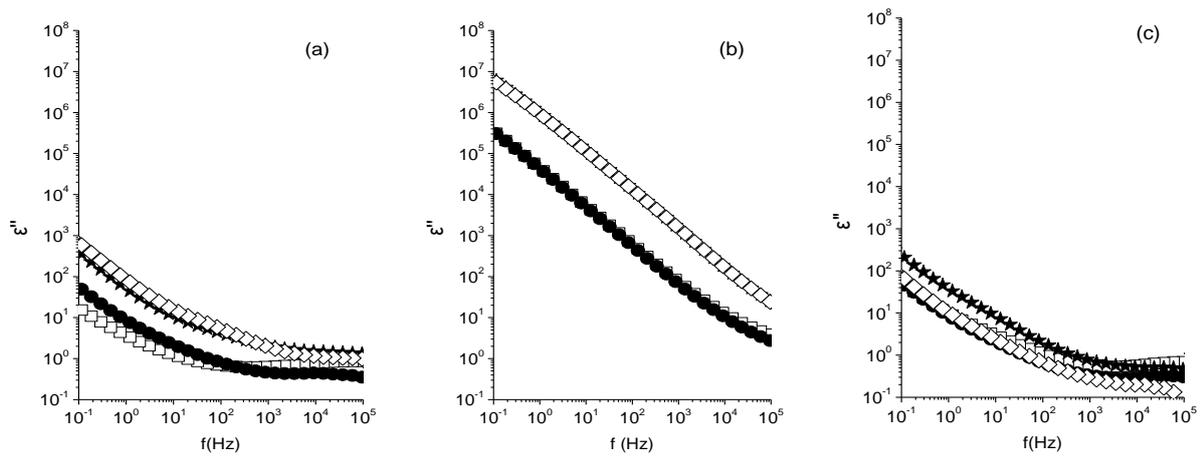

*Figure 1.* Typical isotherms of spectra recorded at different temperatures and pressures for all NGP loading; x=0 % (squares), x=0.05 % (circles), x=0.1 % (stars), x=0.3 % w-w (diamonds)[(a)T=293K  P=1bar, (b)T=353K P=1bar, (c)T=353K P=2000bar]

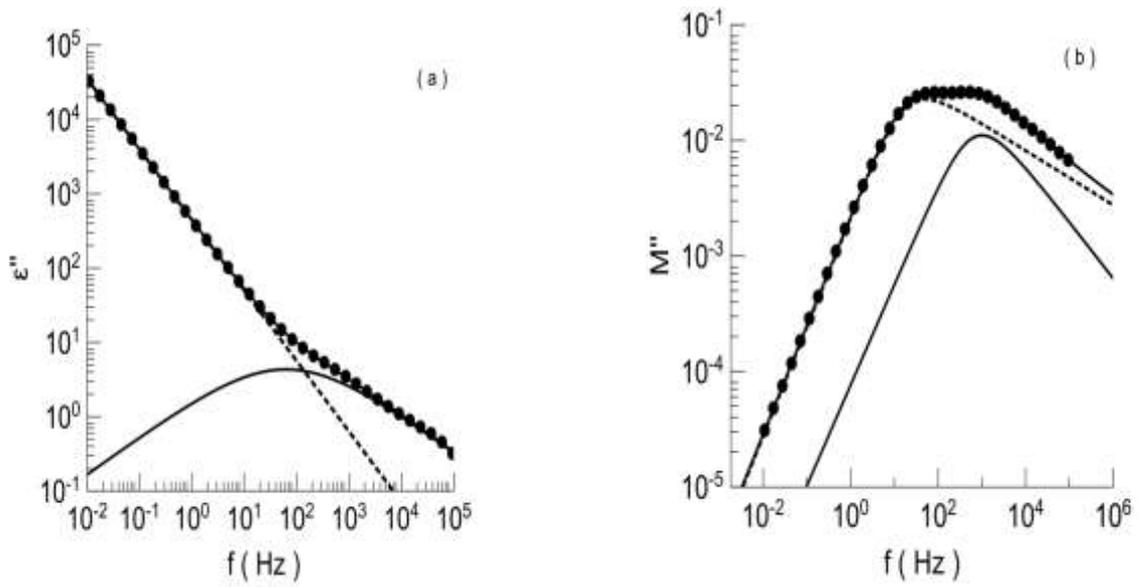

*Figure 2.* Decomposition of the dielectric spectra of (PVA/PVP):(1/1). (0.3 % w-w NGP) recorded at T=373 K and P=1500 bar by fitting eq. (1) (solid line) to the experimental data points. Dash straight line and peak consist the dc-component and the β-relaxation, respectively. The best fitting parameters of eq. (1) are: $log[\sigma_{dc}(S/cm)]= -9.646(5)$  $n= 0.950(1)$, $\Delta\varepsilon= 20.4(3)$, $log[f_0(Hz)]= 1.797(5)$, $a= 0.51(1)$, and $b=1$.

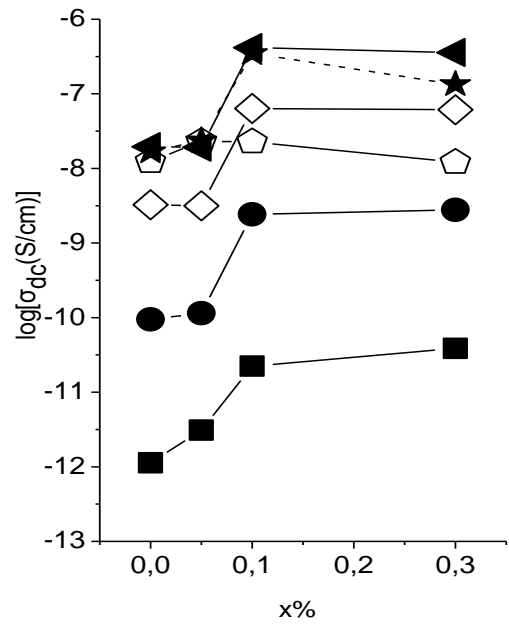

*Figure 3.* The logarithm of the dc-conductivity against NGP mass fraction at P=1 bar. T=293 K (squares), T=313 K (circles), T=333K (diamonds), T=353 K (triangles), T=373 K (stars), T=393 K (polygons).

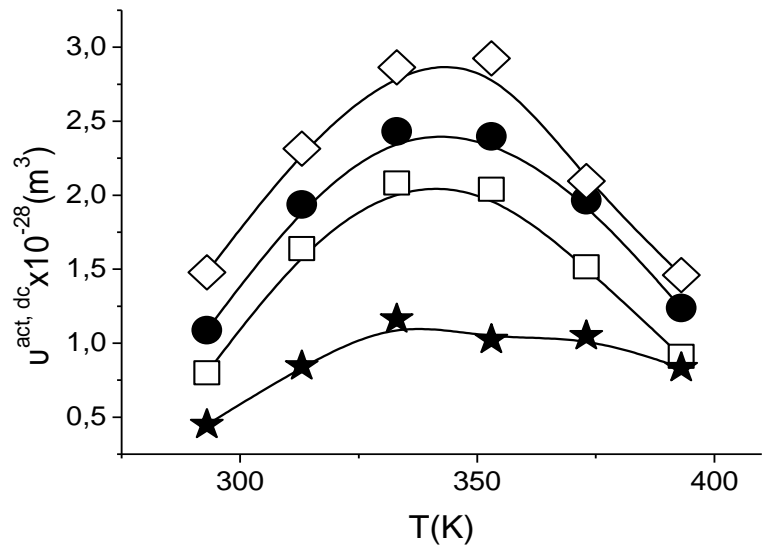

*Figure 4.* Activation volume $v^{act,dc}$ is plotted against temperature for (PVA/PVP):(1/1). and its composites. $x=0$ % (squares), $x=0.05$ % (circles), $x=0.1$ % (stars), $x=0.3$ % w-w (diamonds).

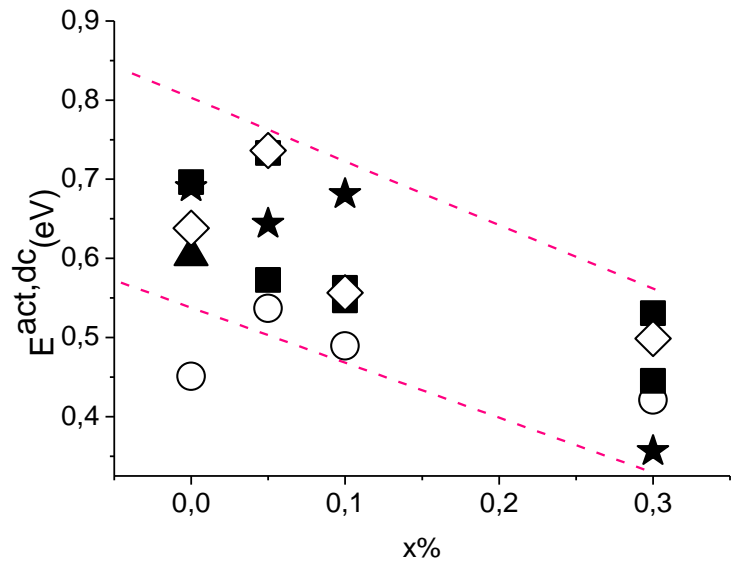

*Figure 5.* *Apparent activation energy for dc-conductivity emerged from the conductivity peak of the M´´(f) spectra, for different NGP loading, against pressure (squares: 1 bar; circles: 500 bar; triangles: 1000 bar; diamonds: 1500 bar; stars:2000 bar). Dash lines have been drawn to improve visualization.*

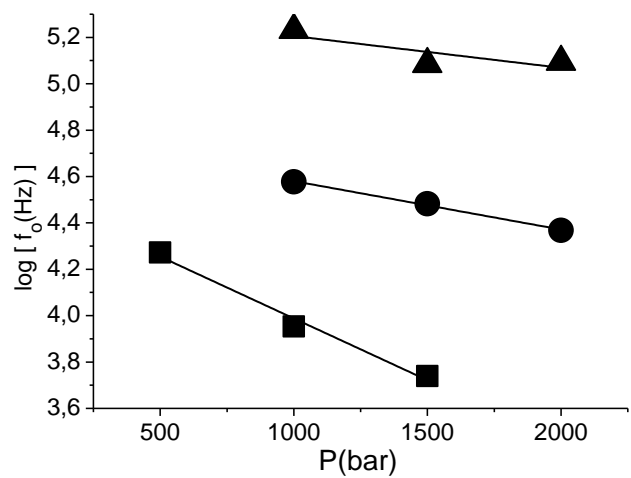

*Figure 6.* Pressure dependence of the maximum of the β-relaxation at different temperatures: T=293 K (squares), T=313 K (circles), T=333 K (triangles).

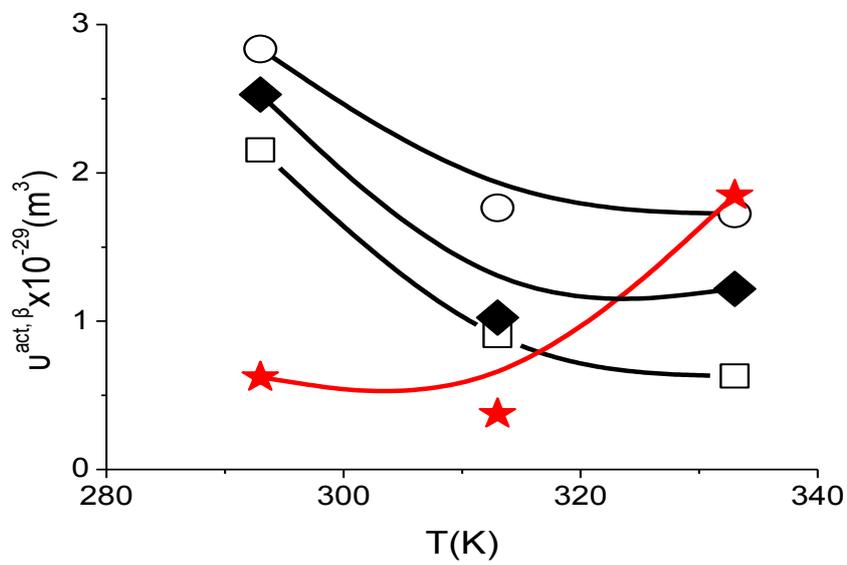

*Figure 7.* Activation volume $v^{act,\beta}$ plotted against temperature for (PVA/PVP):(1/1). and its composites. $x=0$ % (squares), $x=0.05$ % (circles), $x=0.1$ % (stars), $x=0.3$ % w-w (rhombuses)

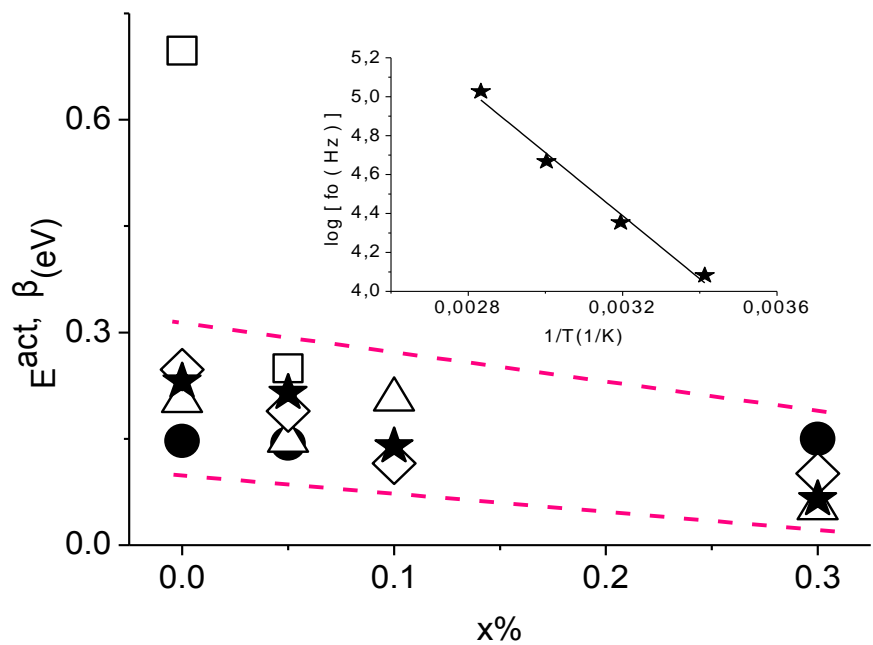

*Figure 8. The combined pressure activation energy for the β-relaxation as a function of w-w % NGP loading, for different pressures (squares: 1 bar; circles: 500 bar; triangles: 1000 bar; diamonds: 1500 bar; stars:2000 bar). Dash lines have been drawn for better visualization. Inset: typical dependence of $f_0$ versus reciprocal absolute temperature for PVA(PVA/PVP):(1/1): 0.1 w-w % NGP. The straight line is the best fit of Arrhenius law to the data points.*